\documentclass[aps,prb,superscriptaddress,reprint,showpacs]{revtex4-1}

\usepackage{hyperref}
\usepackage{units}
\usepackage{upgreek}
\usepackage{graphicx}
\usepackage{amsmath}
\usepackage{color}
\usepackage{verbatim}
\begin{document}
\title{Kerr coefficients of plasma resonances in Josephson junction chains}
\date{\today}
\author{T. Wei\ss l}
\email{weissl@kth.se}
\affiliation{Institut N{\'e}el, CNRS et Universit{\'e} Joseph Fourier, BP 166, F-38042 Grenoble Cedex 9, France}

\author{B. K\"ung}
\affiliation{Institut N{\'e}el, CNRS et Universit{\'e} Joseph Fourier, BP 166, F-38042 Grenoble Cedex 9, France}
\author{E. Dumur}
\affiliation{Institut N{\'e}el, CNRS et Universit{\'e} Joseph Fourier, BP 166, F-38042 Grenoble Cedex 9, France}
\author{A. K. Feofanov}
\affiliation{Institut N{\'e}el, CNRS et Universit{\'e} Joseph Fourier, BP 166, F-38042 Grenoble Cedex 9, France}
\affiliation{Ecole Polytechnique F\'ed\'erale de Lausanne, CH-1015 Lausanne, Switzerland}
\author{I. Matei}
\affiliation{Institut N{\'e}el, CNRS et Universit{\'e} Joseph Fourier, BP 166, F-38042 Grenoble Cedex 9, France}
\author{C. Naud}
\affiliation{Institut N{\'e}el, CNRS et Universit{\'e} Joseph Fourier, BP 166, F-38042 Grenoble Cedex 9, France}
\author{O. Buisson}
\affiliation{Institut N{\'e}el, CNRS et Universit{\'e} Joseph Fourier, BP 166, F-38042 Grenoble Cedex 9, France}
\author{F. W. J. Hekking}
\affiliation{Universit{\'e} Grenoble 1/CNRS, LPMMC UMR 5493, B.P. 166, F-38042 Grenoble, France}
\author{W. Guichard}
\affiliation{Institut N{\'e}el, CNRS et Universit{\'e} Joseph Fourier, BP 166, F-38042 Grenoble Cedex 9, France}
\begin{abstract}
We present an experimental and theoretical analysis of the self- and cross-Kerr effect of extended plasma resonances in Josephson junction chains. We calculate the Kerr coefficients by deriving and diagonalizing the Hamiltonian of a linear circuit model for the chain and then adding the Josephson non-linearity as a perturbation. The calculated Kerr-coefficients are compared with the measurement data of a chain of 200 junctions. The Kerr effect manifests itself as a frequency shift that depends linearly on the number of photons in a resonant mode. By changing the input power on a low signal level, we are able to measure this shift.  The photon number is calibrated using the self-Kerr shift calculated from the sample parameters. We then compare the measured cross-Kerr shift with the theoretical prediction, using the calibrated photon number.
\end{abstract}

\maketitle
\begin{section}{Introduction}

One-dimensional arrays of Josephson junctions, or Josephson junction chains, have received considerable interest for more than three decades. They were originally introduced as a theoretical model system for the study of the zero-temperature superconductor-insulator transition in superconducting granular films~\cite{Efetov1981,Bradley1984}. Josephson junction chains consist of superconducting islands with a small capacitance to ground $C_0$, connected to each other by Josephson junctions. The superconductor-insulator transition originates from the Josephson potential energy $E_J \cos(\phi)$ as this allows the winding by $2 \pi$ of the phase difference $\phi $ between neighboring islands. At zero temperature, these phase-slips are driven by the quantum fluctuations of the phase induced by Coulomb charging effects. As a result, the transition occurs as a function of the parameter $E_J/E_0$, where $E_0 = e^2/2C_0$ the Coulomb charging energy of the island. Since the quantum fluctuations of the phase actually correspond to the zero-point fluctuations of propagating electromagnetic modes along the chain\cite{Masluk2012}, the transition can also be characterized in terms of the impedance $Z/R_Q=\sqrt{E_0/E_J} $ associated with the propagation of these modes, where $R_Q$ is the superconducting resistance quantum $R_Q=h/4e^2$. Indeed, the transition to the insulating state occurs when $Z/R_Q$ becomes larger than $\sim \pi/2$. The superconductor-insulator transition has been observed in granular films~\cite{Jaeger1989} and wires~\cite{Giordano1988}, as well as in  Josephson junction chains~\cite{Chow1998,Haviland2000}.

More recent work focused on Josephson junction chains of finite length for which the junction capacitance $C$ is much larger than the capacitance $C_0$ to ground. In such chains, the above mentioned $2\pi$ phase windings occur in the form of coherent quantum phase-slips (QPSs). It has been shown \cite{Matveev2002,Guichard2010,Pop2010} that coherent QPSs occurring locally at each junction of the chain induce a nonlinearity related to the global charge $q$ on the chain $U_0 \cos(\pi q/e)$, dual to the usual Josephson nonlinearity. Here, the amplitude $U_0 = N u_0$ scales with the length $N$ of the chain. The amplitude $u_0 \simeq (E_J^3 E_C)^{1/4} \exp{-\sqrt{8 E_J/E_C}}$ is the amplitude for a single QPS to occur on one of the junctions of the chain, where $E_C = e^2/2C$. In this limit, the chain behaves as a so-called quantum phase-slip junction (QPSJ), a device dual to the usual Josephson junction and whose properties have been discussed since the pioneering work by Likharev and coworkers~\cite{Likharev1985a,Likharev1985b}. For instance,  a QPSJ  is expected to sustain Bloch oscillations, dual to Josephson oscillations.  Evidence of Bloch oscillations has been found in Refs.~\onlinecite{Kuzmin1991,Watanabe2001,Corlevi2006}. The actual realization of a QPSJ could have far-reaching consequences in quantum metrology and quantum information processing~\cite{Mooij2006,Rastelli2015,Viola2015}.

The effect of the propagating modes of the chain on the properties of the QPSJ has not been taken into account in Refs.~\onlinecite{Matveev2002,Guichard2010,Pop2010}. This is correct for relatively short chains, with $N < 2 C/C_0$. Yet for longer chains, these modes appear and their effect on the phase-slip amplitude is non-negligible~\cite{Rastelli2013,Weissl2015}. Indeed, in the thermodynamic limit, it is the interplay between the modes and the phase-slips that leads to the superconductor-insulator transition. In view of possible QPSJ applications, it is therefore important to study the modes directly, in the presence of the chain's nonlinearity. This is the subject of the present paper, where we present two-tone spectroscopy measurements of the modes for a chain containing 200 junctions and tuned in the weakly nonlinear regime. We show that the results can be interpreted in terms of a model Hamiltonian that takes the weak nonlinearity into account via self-Kerr and cross-Kerr interaction terms of the propagating modes.

The paper is organized as follows. In Section \ref{sec:Experiment}, we introduce the sample and the experimental setup and show spectroscopy measurements of the harmonic modes of a chain of 200 junctions. In Section \ref{sec:Modes}, we derive the Hamiltonian of a Josephson junction chain and its eigenvalues and eigenvectors in the linear limit. We also introduce the lowest order nonlinear terms as a perturbation to the linear Hamiltonian. The two most prominent terms are the self-Kerr and cross-Kerr terms that appear. In Section \ref{sec:nonlinear}, we first use the measured self-Kerr frequency shift to calibrate the photon numbers in the resonant modes of the Josephson junction chain. We then use this calibration to measure the cross-Kerr shifts. We find cross-Kerr shifts of the same order as the ones predicted by theory. In the last section, we summarize our results and conclude.
\end{section}

\section{Experiment}
\label{sec:Experiment}

\subsection{Sample}

\begin{figure}[h]
\center
\includegraphics[width=.9\columnwidth]{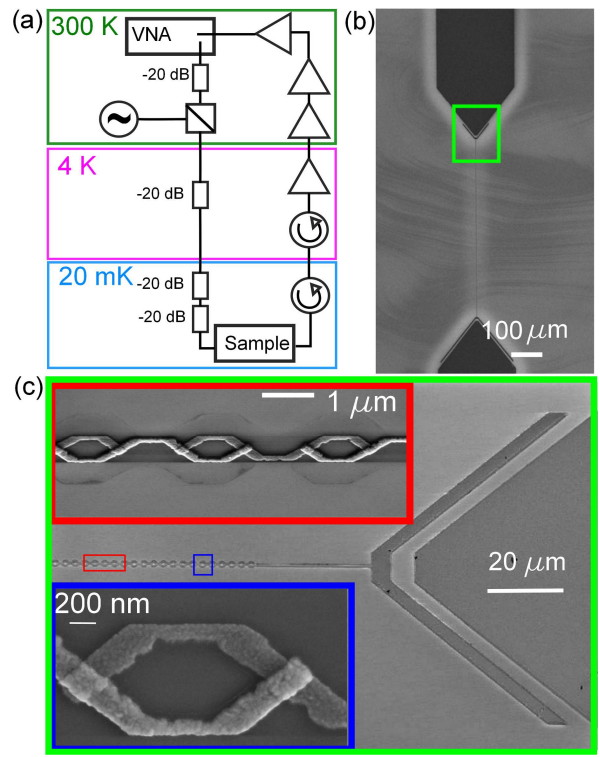}
\caption{(a) Schematic of the measurement setup. (b) SEM-image of the sample. On the top and bottom side one can see the ends of the microstrip transmission lines. The SQUID chain in between is barely visible on this scale. (c) Zoom on one of the coupling capacitors. The image is rotated $90^\circ$ with respect to the top image. The insets show zoom on the chain and on a single SQUID.}
\label{fig:sample}
\end{figure}

We have measured the resonances associated with the propagating modes of a Josephson junction chain containing 200 SQUIDs ({\bf S}uperconducting {\bf Qu}antum {\bf I}nterference {\bf D}evice). The sample is cooled to $\unit{20\,}{\rm{mK}}$ in a dilution refrigerator. The resonant modes are measured with a vector network analyzer (VNA). A scheme of the measurement setup is drawn in figure \ref{fig:sample}a. The input line is attenuated by $\unit{-20}{dB}$ at $\unit{4}{\rm K}$ and $\unit{-40}{dB}$ at base temperature. The input bandwidth is limited by the coax-cables which are specified up to $\unit{18\,}{\rm{GHz}}$ but still show transmission up to higher frequencies. The output line of the measurement setup comprises two cryogenic broadband isolators (Pamtech CW1019-K414), a cryogenic amplifier (Caltech Cryo 1-12 SN262D) and three room temperature amplifiers. The output line has a gain of about $\unit{85\,}{\rm{dB}}$ over a frequency band between $\unit{4\,}{\rm{GHz}}$ and  $\unit{12\,}{\rm{GHz}}$.\\
A SEM-image of the sample is shown in figure \ref{fig:sample}b. The chain is coupled to $\unit{50\,}{\rm{\Omega}}$ microstrip transmission lines at both ends through coupling capacitors. In figure \ref{fig:sample}c we show a zoom on the input coupling capacitor. Input and output capacitor have different sizes, so the chain is coupled to the output line stronger than to the input line. Junctions and transmission lines are fabricated in the same fabrication step by shadow evaporation of aluminum on a silicon substrate. We use $\unit{100\,}{\rm{keV}}$ electron beam lithography with an asymmetric undercut in the two-layer resist (PMMA/MAA, PMMA) to deposit all unwanted structures on the resist walls\cite{Weissl,Lecocq2011}.\\
The sample parameters are summarized in Table \ref{tab:Params}. The Josephson energy was determined by measuring the resistance of test junctions on the same chip at room temperature and by assuming an increase of the resistance of $30\%$ during cooldown, which is what is typically observed for our samples. The junction capacitance is determined from the junction size obtained by SEM-observation of the test junctions. We use a capacitance to surface ratio of $\unit{48\,}{\rm{fF/\mu m^2}}$ \cite{Delsing1990,Lichtenberger1989}.
\begin{table}
\center
\begin{tabular}{c|c|c|c|c}
\hline\hline
$E_J$ (GHz) & $E_C$ (GHz)& $\frac{\omega_p}{2\pi}$ (GHz)  & C (fF) & $L_J$ (nH)\\
\hline
$40.1 \pm 2$ & $ 4.7 \pm 0.5$ & $38.4 \pm 3$ & $4.2 \pm 0.4$ & $4.1 \pm 0.2$ \\
\hline\hline
 $I_c$ (nA) & $C_{in} $(fF) & $C_{out} $(fF) & $C/C_0$  & $C_0$ (aF)   \\
\hline
$81 \pm 4$&$ 8 \pm 2 $ & $ 24  \pm 6$ & $ 35\pm 3 $ & $120 \pm 15$   \\
\hline\hline
\end{tabular}
\caption{Device parameters of the sample.}
\label{tab:Params}
\end{table}

\subsection{Measurements}
Three of the modes of the chain lie within the bandwidth of the amplification chain and can be observed directly by applying a single microwave tone and measuring the transmission coefficient $S_{21}$ between ports 1 and 2 of the VNA. The frequencies of the other modes lie outside the measurement bandwidth.
 
\begin{figure}
\includegraphics[width=\columnwidth]{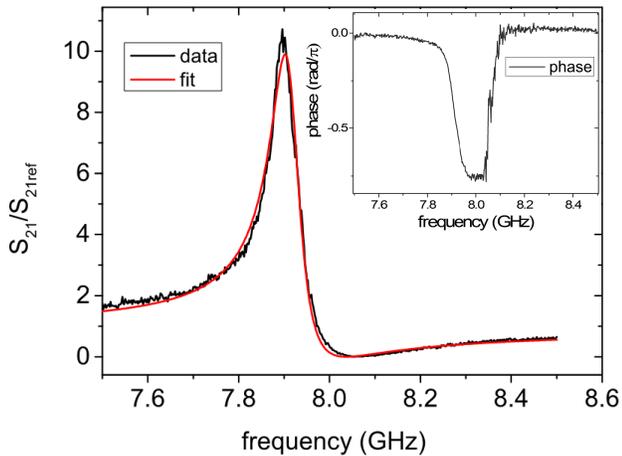}
\caption{Transmitted amplitude through the j=3 resonance. The red curve shows a fit with the Fano formula eq. (\ref{eq:Fano}) using the following fit parameters: $y_0=0.89$, $q=-3.17$, $\gamma=\unit{77.3}{\, \rm{MHz}}$ and $f_0=\unit{7.91}{\, \rm{GHz}}$. The inset shows the transmitted phase as a function of frequency for the same resonance.}
\label{fig:res}
\end{figure}

\begin{figure}[h]
\includegraphics[width=\columnwidth]{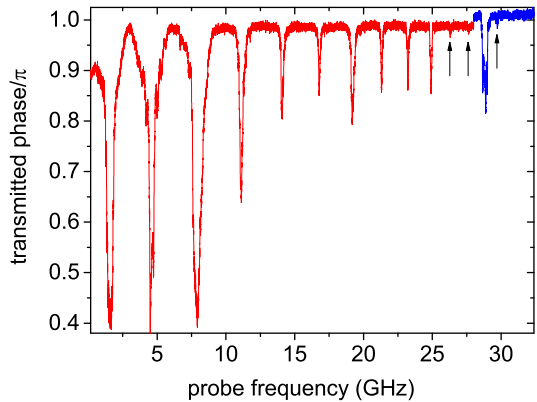}
\caption{Two-tone spectroscopy: phase of the transmission through the $j=2$ resonance as a function of the frequency of an additional probe tone applied by an external microwave source. We observe the lowest 14 resonant modes of the chain. The blue datapoints at higher frequencies were measured with higher excitation power of the probe tone.}
\label{fig:2ton}
\end{figure}

\begin{figure}[h]
\includegraphics[width=\columnwidth]{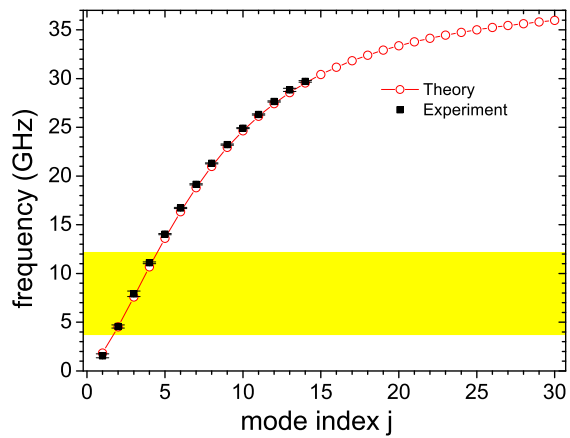}
\caption{Measured frequencies (extracted from the measurement shown in figure \ref{fig:2ton}) of the resonances of a chain of 200 SQUIDs compared to the theoretical prediction. The values of the ground capacitance as well as the coupling capacitors were used as fit parameters in order to find best agreement with the data. The highlighted area corresponds to the band in which a direct spectroscopy is possible with our measurement setup.}
\label{fig:disp}
\end{figure}

As an example for the direct spectroscopy of resonances within the experimental accessible bandwidth, we plot the transmission magnitude $S_{21}$ in figure~\ref{fig:res} for the mode corresponding to $j=3$. The spectrum is divided by a reference spectrum recorded at half flux frustration of the SQUID loop. In this way, the transmission background is normalized to 1. Due to a parasitic coupling between the two transmission lines (input and output lines) the resonances show a Fano line shape\cite{Weissl,Dotan2012,Miroshnichenko2010}. We thus fit the resonances with the Fano formula\cite{Dotan2012} for the transmitted amplitude $y(f)$,
\begin{equation}
\label{eq:Fano}
y(f) =  y_0 \frac{(q + 2(f-f_0)/\gamma)^2}{1+ 4(f-f_0)^2/\gamma^2}.
\end{equation}
Here $f_0$ is the resonance frequency, $\gamma$ the width of the resonance (full width half maximum FWHM), $y_0$ is the amplitude related with the parasitic coupling between the two transmission lines, and $q$ is the Fano factor that is given by the ratio between the amplitudes transmitted at resonance through the chain and the parasitic transmission.\\
Modes outside the band can nevertheless be observed using two-tone spectroscopy. The transmission through one of the resonances inside the band is measured, while sweeping a second probe tone that is applied by an additional microwave source (Agilent 8257D) through a power combiner (Mini-Circuits ZFRSC-183-S). This detection method is based on the nonlinear interaction between the modes.
We plot the phase of the transmitted signal through the $j=2$ resonance as a function of the probe frequency in Fig.\ref{fig:2ton}. We observed the first 14 modes out of the 199 predicted modes of the chain. \\
In figure~\ref{fig:disp}, we show the dispersion relation (frequency as a function of the mode index) extracted from the two-tone spectroscopy in figure~\ref{fig:2ton}.
As will be shown in detail below, the observed frequencies are in good agreement with the theoretical prediction.

\section{Theory}
\label{sec:Modes}

\begin{figure}[h]
\includegraphics[width=0.8\columnwidth]{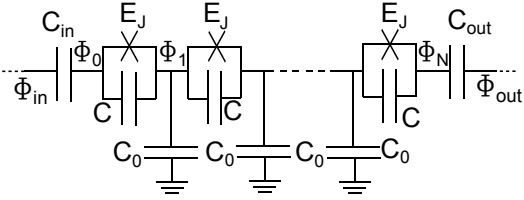}
\caption{Circuit diagram considered for the derivation of the Hamiltonian. A series array of Josephson junctions with Josephson energy $E_J$, capacitance $C$ and ground capacitance $C_0$ coupled to the external circuitry via the capacitors $C_{in}$ and $C_{out}$.}
\label{fig:figure1}
\end{figure}

In this section, we will analyze our experiments based on the circuit model depicted in figure~\ref{fig:figure1}. It consists of a chain of $N$ junctions (the SQUIDs are modeled as single junctions) with Josephson energy $E_J$ and capacitance $C$. The ground capacitance of the superconducting islands is $C_0$. The chain is coupled to the outside world via input and output capacitors $C_{in}$ and $C_{out}$ at the two ends of the chain. The corresponding Lagrangian reads

\begin{align}
\label{eq:Lagmode0}
{\mathcal L}=&\sum_{n=1}^{N-1} \left(\frac{\dot{\Phi}_n^2 C_0}{2} \right)+ \frac{C_{in}}{2} \dot{\Phi}_0^2 +  \frac{C_{out}}{2} \dot{\Phi}_N^2 \nonumber\\
+&  \sum_{n=0}^{N-1} \frac{1}{2}(\dot{\Phi}_{n+1} - \dot{\Phi}_{n})^2 C - \sum_{n=0}^{N-1} E_J \cos(\phi_{n+1} - \phi_{n}),
\end{align}
where we denoted the superconducting phase of the $n$-th island as $\phi_n$, and the corresponding node flux $\Phi_n = \hbar \phi_n /2e$. In the absence of a voltage bias, the input and output capacitors are grounded; we then set $\Phi_{in}=\Phi_{out}=0$, the reference phase for ground being chosen equal to zero. Below, we will first analyze Lagrangian~(\ref{eq:Lagmode0}) by linearizing the Josephson term, i.e. expanding it up to the second order in $(\phi_{n+1} - \phi_n)^2$. Then we will study the effect of weak nonlinearity perturbatively by including fourth order corrections $\propto (\phi_{n+1} - \phi_n)^4$.

\subsection{Linear modes}

In the linear limit, Lagrangian~(\ref{eq:Lagmode0}) of the system can easily be written down as a sum over the islands of quadratic terms only
\begin{align}
\label{eq:Lagmode1}
{\mathcal L}=&\sum_{n=1}^{N-1} \left(\frac{\dot{\Phi}_n^2 C_0}{2} \right)+ \frac{C_{in}}{2} \dot{\Phi}_0^2 +  \frac{C_{out}}{2} \dot{\Phi}_N^2 \nonumber\\
+&  \sum_{n=0}^{N-1} \frac{1}{2}(\dot{\Phi}_{n+1} - \dot{\Phi}_{n})^2 C - \sum_{n=0}^{N-1} \frac{1}{2L}  (\Phi_{n+1} - \Phi_{n})^2\nonumber\\
=& \frac{1}{2}  \vec{\dot{\Phi}}^T  \widehat{C}  \vec{\dot{\Phi}} -  \frac{1}{2}  \vec{\Phi}^T  \widehat{L}^{-1} \vec{\Phi},
\end{align}
where we introduced the Josephson inductance $L = (\hbar/2e)^2/E_J$. In the last line, we introduced the matrix form with the flux vector $\vec{{\Phi}}$ and its transpose $\vec{{\Phi}}^T$ defined as $\vec{{\Phi}}^T  = \left({\Phi}_0, {\Phi}_1,\hdots , {\Phi}_N\right) $ and the capacitance matrix $\widehat{C}$ and the inverse inductance matrix $\widehat{L}^{-1}$.

From Lagrangian (\ref{eq:Lagmode1}) the Hamiltonian can be derived via Legendre-transformation
\begin{equation}
\begin{split}
H_0=& \vec{Q}^T \vec{\dot{\Phi}} - {\mathcal L}
= \frac{1}{2}  \vec{Q}^T  \widehat{C}^{-1}\vec{Q} + \frac{1}{2}   \vec{\Phi}^T  \widehat{L}^{-1}  \vec{\Phi}.
\label{eq:modHam}
\end{split}
\end{equation}
where the components of the charge vector are defined by $ Q_n= \frac{\partial {\mathcal L}}{ \partial \dot{\Phi}_n}$.

The Hamiltonian  (\ref{eq:modHam}) can be quantized and expressed in terms of the usual bosonic creation and annihilation operators. Details are presented in appendix~\ref{App:A}. The Hamiltonian \ref{eq:modHam} then reads
\begin{equation}
\hat{H}_0=\sum_{j=1}^N \hbar \omega_j \left(\hat{a}_j^\dag\hat{a}_j+\frac{1}{2}\right),
\label{eq:Hamq}
\end{equation}
where the eigenfrequencies are given by the eigenvalue problem
\begin{equation}
\widehat{C}^{-1/2} \widehat{L}^{-1} \widehat{C}^{-1/2}  \vec{\psi_{j}} = \omega_j^2  \vec{\psi_{j}},
\label{eq:eigp}
\end{equation}
with $ \widehat{C}^{-1/2}  \widehat{C}^{-1/2} = \widehat{C}^{-1}$. The eigenvectors $ \vec{\psi_{j}} $ of the matrix $\widehat{C}^{-1/2} \widehat{L}^{-1} \widehat{C}^{-1/2} $ are related with the eigenmodes of the Hamiltonian (\ref{eq:modHam}) via
\begin{equation}
\vec{\hat{\Phi}}=  \sum_{j}  \widehat{C}^{-1/2}\vec{\psi}_{j} \sqrt{\frac{\hbar}{2\omega_j}} \left( \hat{a}_j + \hat{a}_j^\dag\right) . \label{eq:Fluxop}
\end{equation}

\begin{figure}[h]
\includegraphics[width=\columnwidth]{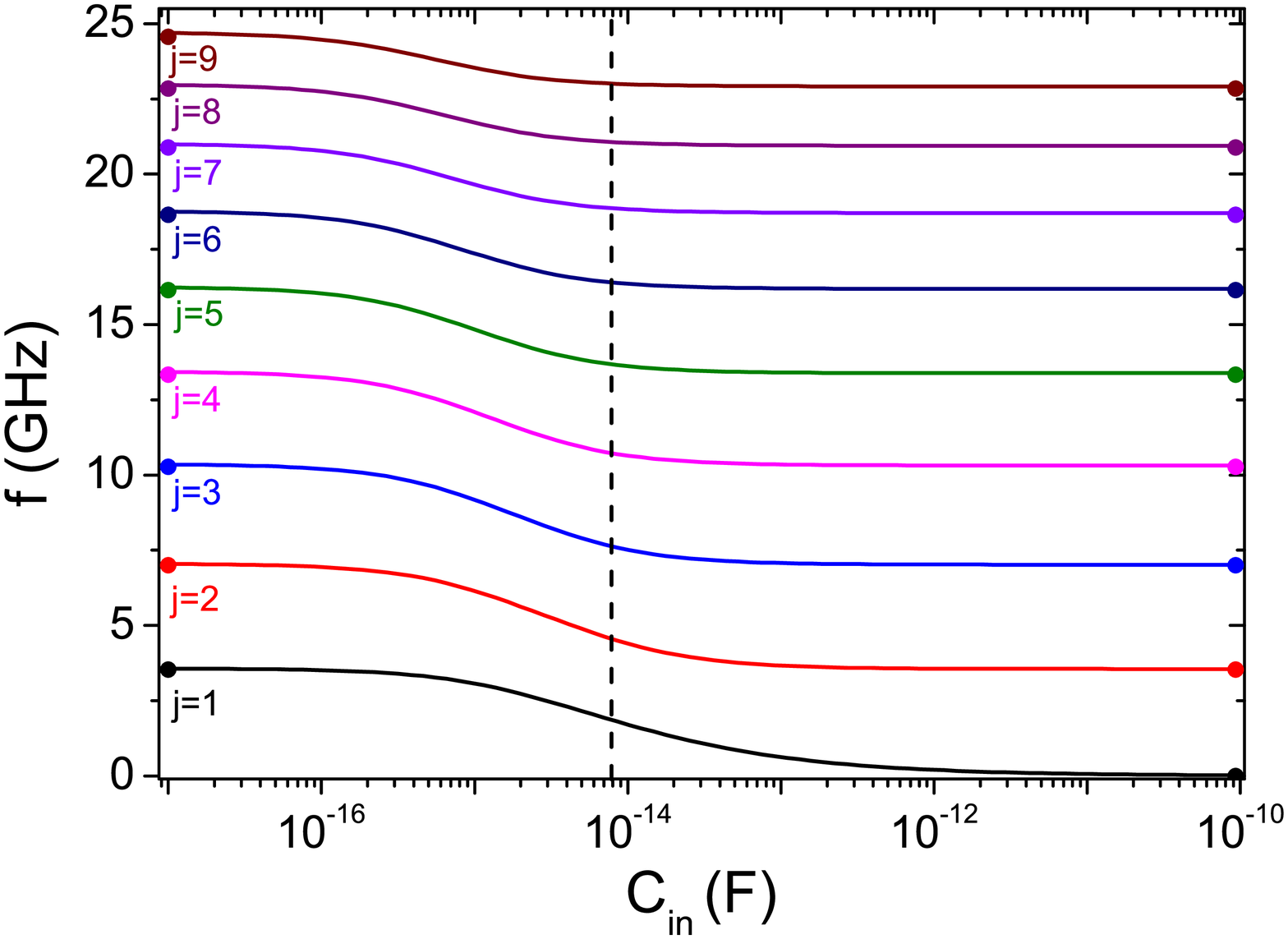}
\caption{Eigenfrequencies as a function of the coupling capacitor. We fixed $C_{out}=3C_{in}$. All other parameters were taken from the sample parameters listed in Table \ref{tab:Params}. The eigenvalues were calculated using Eq. (\ref{eq:eigp}). The black vertical line indicates the coupling capacitances realized in the experiment. The solid dots indicate the analytical values for small ($C_{in}=0$) and and large ($C_{in} \to \infty $) coupling capacitors. }
\label{fig:foC}
\end{figure}

\begin{figure}
\includegraphics[width=\columnwidth]{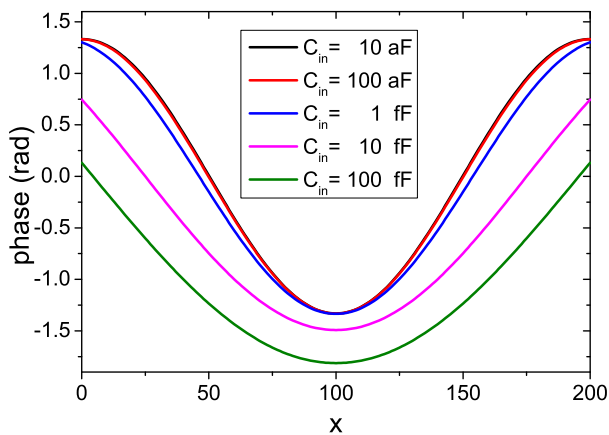}
\caption{Space dependent part of the eigenfunction for the j=2 mode for different values of the coupling capacitance $C_{in}$. ($C_{out}=3C_{in}$) }
\label{fig:eigf}
\end{figure}

The eigenvalues $\omega_j$ of the Hamiltonian (\ref{eq:Hamq}) are very sensitive to the values of the coupling capacitors $C_{in}$ and $C_{out}$. This is because these capacitances determine the boundary conditions for the allowed eigenfunctions. To illustrate this, we plot the eigenfrequencies of the lowest modes as a function of the input capacitance in figure \ref{fig:foC}. We keep a fixed ratio between output and input capacitors; the other parameters were chosen according to the sample parameters listed in Table \ref{tab:Params}. We see that, upon increasing $C_{in}$, the eigenfrequencies decrease monotonically. Indeed, when $C_{in} = C_{out}= 0$, the chain is effectively isolated; as a result the allowed eigenfunctions satisfy the zero-current boundary condition $\Phi_0 - \Phi_1 = \Phi_{N-1} - \Phi_N = 0$. The corresponding $N-1$ allowed modes are given by $\Phi_n = A_k \cos [k(n-1/2)]$, where the constant $A_k$ fixes the normalization and the dimensionless wave vector $k = j\pi/(N-1)$ with $j=0,1, \ldots, N-2$. In the opposite limit of large $C_{in}$, and hence $C_{out}$, islands $0$ and $N$ are perfectly coupled to the outside electrodes connected to ground, thus the boundary condition reads $\Phi_0 = \Phi_N = 0$. The corresponding $N-1$ modes are now given by $\Phi_n = B_k \sin (k n)$, where the constant $B_k$ fixes again the normalization and the dimensionless wave vector $k = j\pi/N$ with $j=1,2, \ldots, N-1$. The modes frequencies follow from the usual dispersion relation, valid for homogeneous chains\cite{Masluk2012}
\begin{equation}
\omega_k = \omega_p \sqrt{\frac{1 - \cos k}{1-\cos k + C_0/2C}}.
\end{equation}
Their limiting values are indicated in Fig.~\ref{fig:foC},
both for $C_{in} = 0 $ and for $C_{in} = \infty$. The numerical solution for finite $C_{in}$ smoothly interpolates between these limiting values. The gradual change of the nature of eigenfunctions as $C_{in}$ is increased is shown in Fig.~\ref{fig:eigf}. One sees that the eigenfunctions indeed have zero slope at the entrance and exit of the chain for small $C_{in}$, whereas they satisfy the zero-phase boundary condition for large $C_{in}$.

When comparing the theory with the experimental data, we have adjusted the values of the ground capacitance $C_0$ and the coupling capacitances $C_{in}$ and $C_{out}$ in order to find best agreement with the experimentally observed resonance frequencies. The result is shown in Fig.~\ref{fig:disp}, where we plot the observed frequencies as a function of the mode index j. The observed frequencies are in good agreement with the theoretical prediction.

\subsection{Weak nonlinearity}
\label{sec:Nonlin}
We now include the weak nonlinearity, namely the quartic term of the Josephson energy $-E_J (\phi_{n+1} - \phi_n)^4/24$ in the Hamiltonian and treat it as a perturbation to the linear Hamiltonian (\ref{eq:modHam})~\cite{Bourassa2012}. The perturbative approach is justified if the mode frequency $\omega_j \ll E_J/\hbar$ which is the case for the low frequency modes of our sample.\\
Using the mode expansion (\ref{eq:Fluxop}), the modified Hamiltonian $\hat{H}_{NL}=\hat{H}_0+\hat{U}_{NL}$ can be found, with a nonlinear contribution $\hat{U}_{NL}$,
\begin{equation}
\begin{split}
\label{eq:NLHam}
\hat{H}_{NL}=& \sum_j \hbar \omega'_j   \hat{a}_j^\dag\hat{a}_j - \sum_j \frac{\hbar}{2} K_{jj}  \hat{a}_j^\dag\hat{a}_j\hat{a}_j^\dag\hat{a}_j \\
&-  \sum_{\begin{array}{c}j,k \\(j\neq k)\end{array}} \frac{\hbar}{2} K_{jk}  \hat{a}_j^\dag\hat{a}_j\hat{a}_k^\dag\hat{a}_k\\
&-\sum_{\begin{array}{c}j,k \\(j\neq k)\end{array}} \frac{\hbar}{2} \zeta_{jjk}  (\hat{a}_j^\dag\hat{a}_j+\frac{1}{2}) (\hat{a}_j^\dag\hat{a}_k +\hat{a}_k^\dag\hat{a}_j ) \\
&-\sum_{\begin{array}{c}j,k,l \\(j\neq k\neq l)\end{array}}   \frac{\hbar}{2} \zeta_{jkl}  (\hat{a}_j^\dag\hat{a}_j+\frac{1}{2}) (\hat{a}_k^\dag\hat{a}_l +\hat{a}_l^\dag\hat{a}_k ). \\
\end{split}
\end{equation}
The coefficient $K_{jj}$ is called self-Kerr coefficient and the corresponding self-Kerr term of the Hamiltonian (\ref{eq:NLHam}) gives a frequency shift of the frequency of the mode $j$ that scales linearly with the photon number $\hat{a}_j^\dag\hat{a}_j$ in the mode $j$. At higher photon number, this term can give rise to photon blockade\cite{Bourassa2012,Imamoglu1997} and bistability of the resonator. The term containing the cross-Kerr coefficient $K_{jk}$ causes a shift of the frequency of the mode $j$ that depends linearly on the photon number in the mode $k$. As will be discussed in more detail below, the cross-Kerr effect can be used to probe the photon numbers. The cross- and self-Kerr coefficients calculated with Eqs.~(\ref{eq:Kjj}) and (\ref{eq:Kjk}) for the parameters of the sample studied in this paper are listed in Table \ref{tab:Kerr}. Finally, the two terms containing the coefficients $\zeta_{jkl}$ give rise to a coupling between two different modes that depends on photon numbers in a third mode. These terms can also give small corrections to the self- and cross-Kerr coefficients in second order. These corrections are small by a factor $K/\Delta$, where K is the relevant Kerr coefficient and $\Delta$ the difference between the mode frequencies involved in the second order process.\\
The nonlinear coefficients are given by
\begin{align}
K_{jj} &= \frac{2\hbar \pi^4 E_J \eta_{jjjj}}{\Phi_0^4 C^2\omega_j^2} \label{eq:Kjj}\\
K_{jk} &= \frac{4\hbar \pi^4 E_J \eta_{jjkk}}{\Phi_0^4 C^2\omega_j\omega_k} \label{eq:Kjk} \\
\zeta_{jkl} &= \frac{4\hbar \pi^4 E_J \eta_{jjkl}}{\Phi_0^4 C^2\omega_j\sqrt{\omega_k\omega_l}}\label{eq:zeta} ,
\end{align}
and the renormalized mode frequency is given by
\begin{equation}
\omega'_j = \omega_j - K_{jj}/2 - \sum_k K_{jk}/4 .
\label{eq:omprim}
\end{equation}
Here we introduced the dimensionless quantities
\begin{equation}
\begin{split}
\eta_{jklp} =\sum_n &\left[ \sum_m \left( \left(\sqrt{C}\widehat{C}_{n,m}^{-1/2}-\sqrt{C}\widehat{C}_{n-1,m}^{-1/2}\right)\psi_{m,j} \right) \right. \\
&\sum_m \left(\left(\sqrt{C}\widehat{C}_{n,m}^{-1/2}-\sqrt{C}\widehat{C}_{n-1,m}^{-1/2}\right)\psi_{m,k}  \right) \\
& \sum_m  \left(\left(\sqrt{C}\widehat{C}_{n,m}^{-1/2}-\sqrt{C}\widehat{C}_{n-1,m}^{-1/2}\right)\psi_{m,l} \right)\\
&\left.\sum_m \left(\left(\sqrt{C}\widehat{C}_{n,m}^{-1/2}-\sqrt{C}\widehat{C}_{n-1,m}^{-1/2}\right)\psi_{m,p}\right)\right]
\label{eq:eta}
\end{split}
\end{equation}
that take into account the spatial variation of the modes.$\psi_{m,j}$ is the m-th component of the vector $\vec{\psi}_j$ defined by Eq. \ref{eq:eigp}.\\
The details of the derivation of Eqs.~(\ref{eq:NLHam}), (\ref{eq:eta}) and (\ref{eq:omprim}) can be found in appendix B.
\begin{table}
\center

\begin{tabular}{|c|c|c|c|}
\hline
mode index & 2 & 3 & 4 \\
\hline
2 & \unit{2.8 $\pm$0.4}{$\,$MHz} & \unit{6.6 $\pm$0.9}{$\,$MHz}& \unit{9.4 $\pm$ 1.4}{$\,$MHz}\\
\hline
3 & \unit{6.6 $\pm$0.9}{$\,$MHz} & \unit{8.3 $\pm$0.7}{$\,$MHz}& \unit{15.9 $\pm$ 2.4}{$\,$MHz}\\
\hline
4 & \unit{9.4 $\pm$1.4}{$\,$MHz}&  \unit{15.9 $\pm$ 2.4 }{$\,$MHz}& \unit{16.7 $\pm$ 2.5}{$\,$MHz}\\
\hline
\end{tabular}
\caption{Kerr coefficients for the modes j=2,3 and 4 predicted by our theory. The errors were estimated assuming the maximum errors for $C$ and $C_0$.}
\label{tab:Kerr}
\end{table}
It is noteworthy that although the expressions for the Kerr coefficients (Eqs: \ref{eq:Kjj},\ref{eq:Kjk},\ref{eq:zeta}) explicitly contain the Josephson energy $E_J$, the Kerr coefficents actually do not depend on $E_J$. Indeed the mode frequencies $\omega_j$ also contain a factor $\sqrt{E_J}$ so that the dependence on the Josephson energy cancels out.
\section{Experiments in the nonlinear regime}
\label{sec:nonlinear}

\subsection{Photon number calibration with the self-Kerr shift}
\label{sec:selfkerr}
The self-Kerr shift can be used to calibrate the photon number in a resonant mode by measuring the resonance frequency as a function of the drive power.
The resonance frequencies as a function of power are obtained by fitting the resonance curves for each excitation power.
\begin{figure}[!h]
\center
\includegraphics[width=.9\columnwidth]{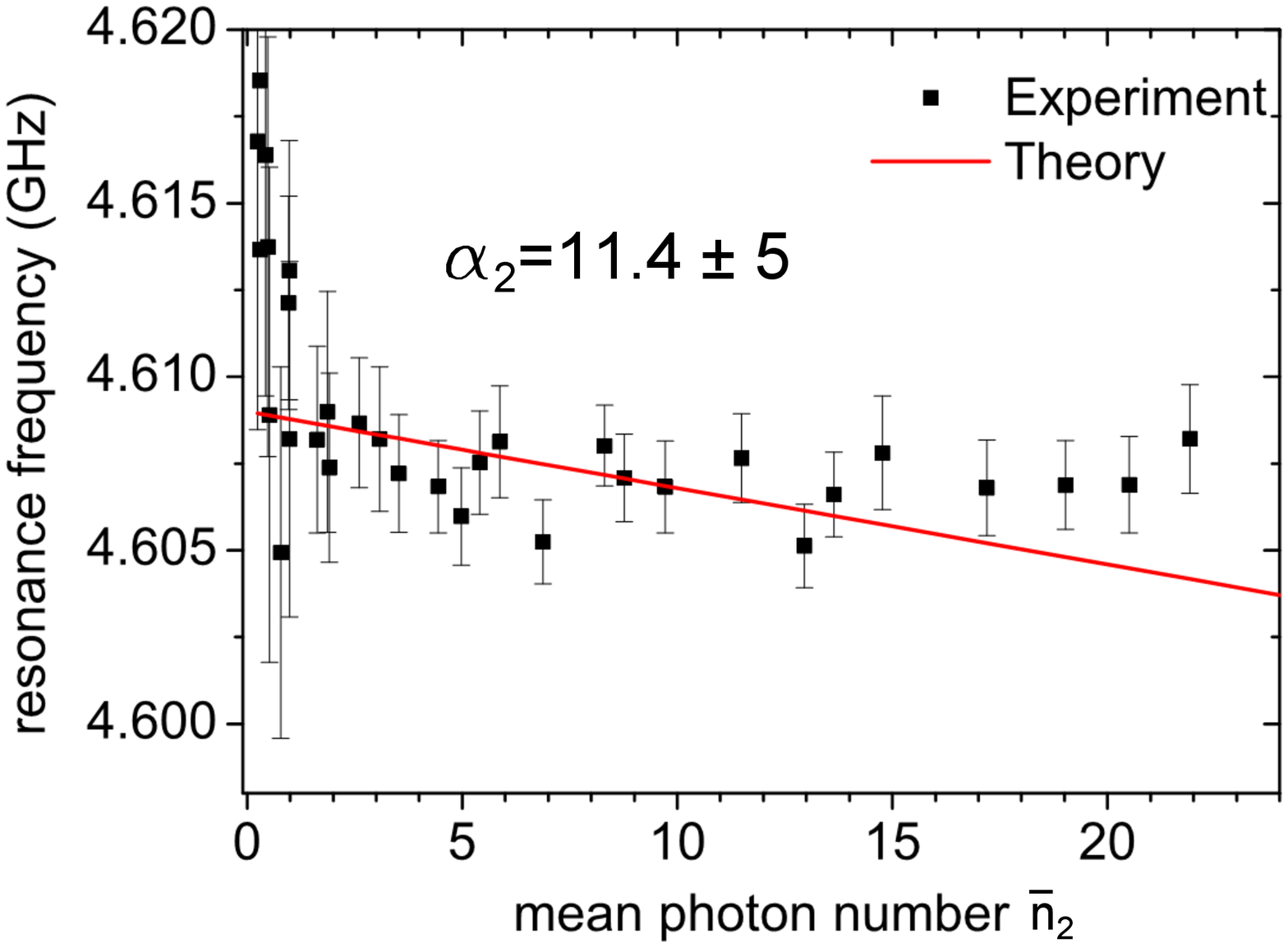}
\includegraphics[width=.9\columnwidth]{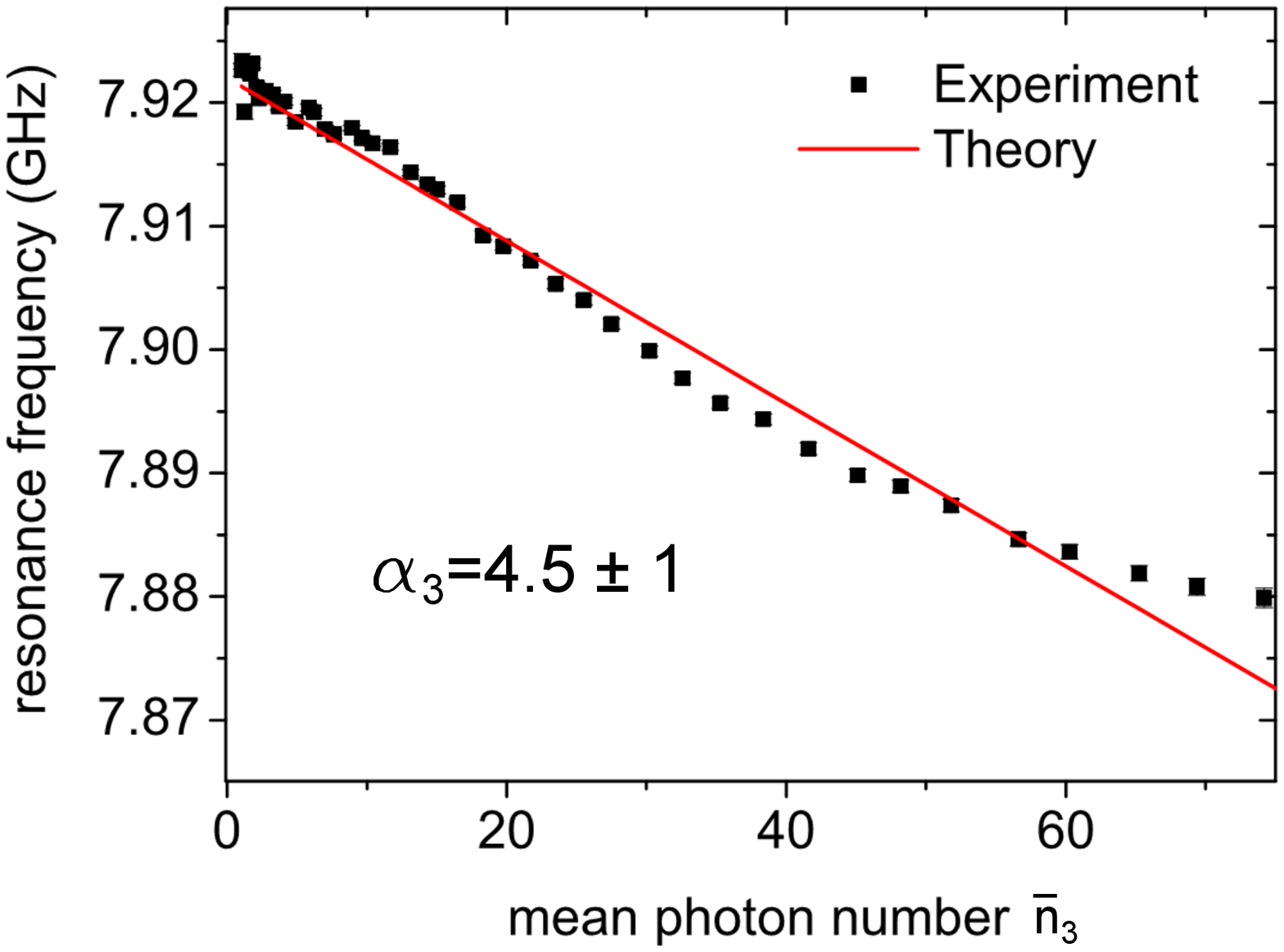}
\includegraphics[width=.9\columnwidth]{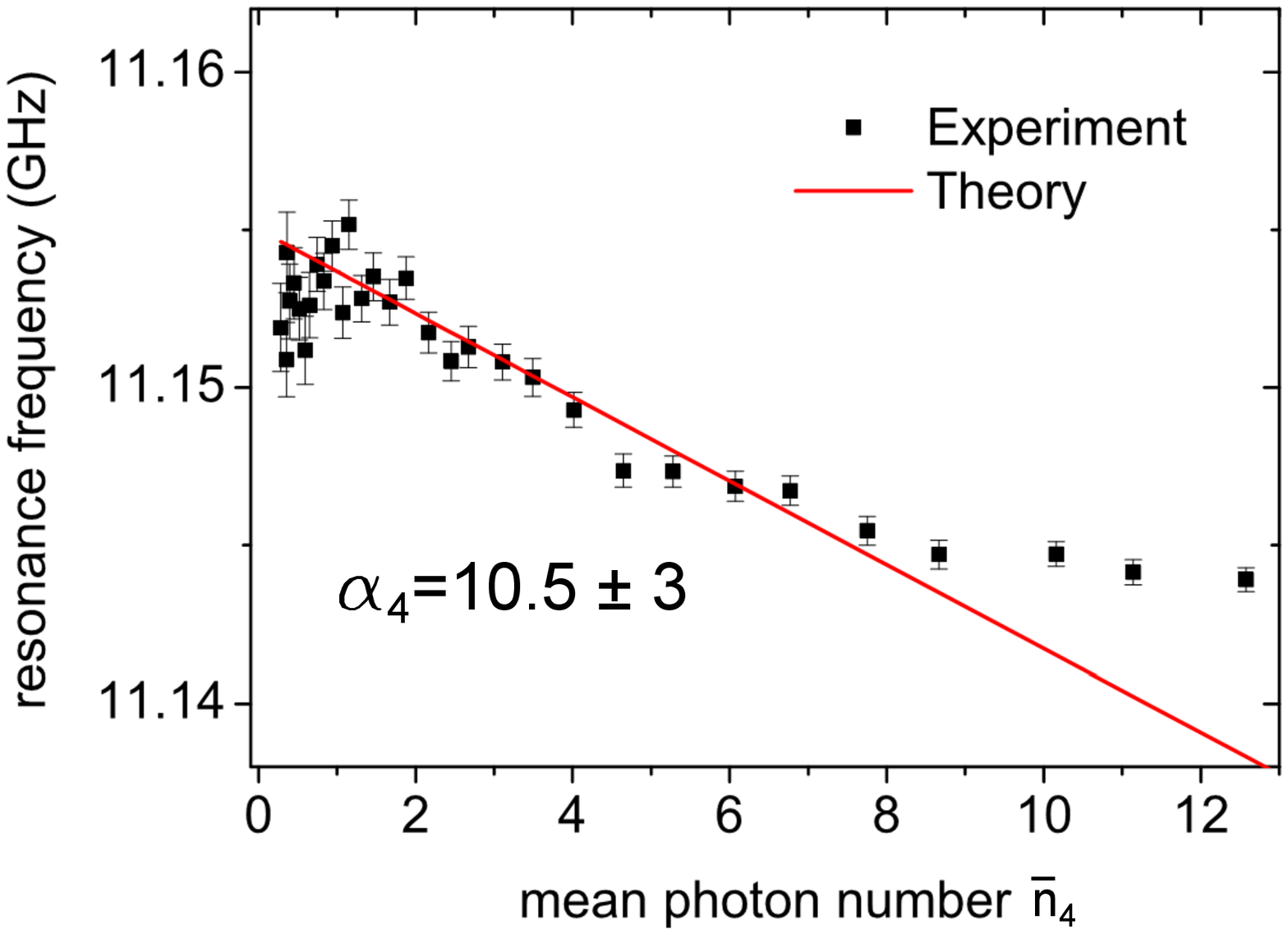}
\caption{Experimental self-Kerr shift compared to theory. We used a factor $\alpha_j$ to adjust the experimental photon number. The three graphs correspond to the three modes of the chain $j=2$ (top), $j=3$ (center) and $j=4$ (bottom) that lie within the bandwidth of the measurement setup.}
\label{fig:1Kerrex}
\end{figure}
If only one mode of the chain is excited, the Hamiltonian Eq.~(\ref{eq:NLHam}) reduces to
\begin{equation}
\hat{H}_{SK}= \left( \hbar \omega'_j   - \frac{\hbar}{2} K_{jj}  \bar{n}_j \right)\hat{a}_j^\dag\hat{a}_j . \\
\end{equation}
The mean photon number $\bar{n}_j$ of a mode is related with the incident power $P_{in}$ via the relation \cite{PalaciosPhD}
\begin{equation}
\bar{n}_j=\frac{4\gamma_{j,in}} {\hbar \omega_j \gamma_{j,tot}^2} P_{in}
\label{eq:nj}
\end{equation}
Here $ \gamma_{j,tot}$ is the total decay rate of the mode related with the width of the resonance, $\gamma_{j,in}$ is the decay rate through the input capacitor, $P_{in}$ is the incident power to the sample. Equation \ref{eq:nj} can be rewritten in terms of the input power at room temperature
\begin{equation}
\bar{n}_j=\frac{4} {\hbar \omega_j \gamma_{j,tot}} A P \frac{1}{\alpha_j}.
\end{equation}
where A is the attenuation of the input line of the cryostat, $P$ is the input power at room temperature and $\alpha_j$ is a numerical factor that takes into account the uncertainty of the attenuation $A$, reflections at the input of the sample as well as the ratio $\gamma_{j,in}/ \gamma_{j,tot}$.

\begin{figure*}[!th]
\center

\includegraphics[width=.75\textwidth]{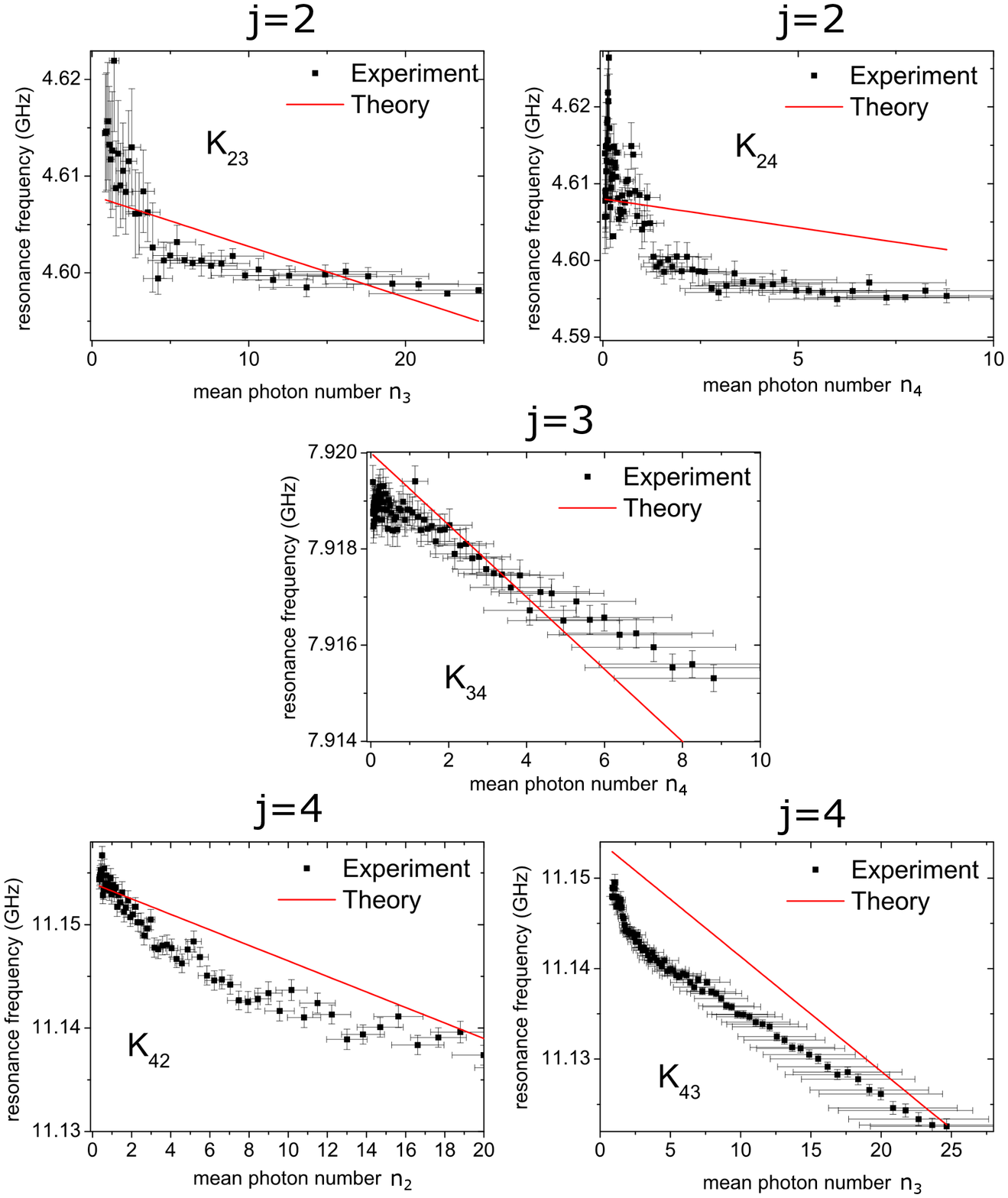}

\caption{Cross-Kerr shift $K_{jk}$ between the $j=2$, $j=3$ and $j=4$ mode as a function of the photon numbers in the mode  $k=2$, $k=3$ and $k=4$. The photon number in the mode $k$ was calibrated with the self-Kerr shift. For two drives the Kerr Hamiltonian takes the form $\hat{H}_{CK}=  \left(\hbar \omega'_j  - \frac{\hbar}{2} K_{jj}  \bar{n}_j- \frac{\hbar}{2} K_{jk}  \bar{n}_k \right) \hat{a}_j^\dag\hat{a}_j $. 
}
\label{fig:cK}
\end{figure*}
We use the self-Kerr measurements to extract the parameter $\alpha_j$ for each of the modes. In figure \ref{fig:1Kerrex} we plot the experimentally observed self-Kerr shifts together with the theoretical prediction.

The values of $\alpha_j$ that were used to adjust the photon numbers in the modes are shown in the figures.
The non-uniform variation of $\alpha_j$ with the mode index is related with spurious resonances in the sample holder. Indeed the transmission background (off resonance) is highest for the  $j=3$ resonance and smallest for the $j=2$ resonance.\\
In the low power region (the few photon limit) the line shape of the resonance is altered very little by the presence of the non-linearity so that the fitting with the Fano formula Eq.(\ref{eq:Fano}) is possible. In the limit of high drive powers the fitting with the linear model Eq.(\ref{eq:Fano}) becomes less accurate. The vertical error bars in Fig:\ref{fig:1Kerrex}and Fig:\ref{fig:cK} represent the error of the fit to the resonance curve.\\
The perturbation approach breaks down at higher photon number as the terms $K_{jj} n_j$ or $K_{jk} n_k$ are no longer small compared to the mode frequency $\omega_j$.\\
For the high excitation amplitudes the power dependent frequency shift is reversed so that an increase of the resonance frequency with drive power is observed. The onset of this upwards shift is nevertheless visible in figure \ref{fig:1Kerrex} at high photon numbers.  \\
The errors on the photon number calibration ($\alpha_j$) dominate over the uncertainty of the theoretical prediction caused by the uncertainty of the sample parameters (Table: \ref{tab:Kerr}). $\alpha_j$ depends on the range of drive powers that is used for the calibration. We estimated the errors of $\alpha_j$ by performing the photon number calibration for different power ranges.

\subsection{Measurement of the cross-Kerr effect}
\label{sec:crosskerr}

With the photon number calibrated we are now able to measure the cross-Kerr shifts.
For two-tone driving the Hamiltonian  Eq.~(\ref{eq:NLHam}) now takes the form
\begin{equation}
\hat{H}_{CK}=  \left(\hbar \omega'_j  - \frac{\hbar}{2} K_{jj}  \bar{n}_j- \frac{\hbar}{2} K_{jk}  \bar{n}_k \right) \hat{a}_j^\dag\hat{a}_j .
\end{equation}
To experimentally investigate the cross-Kerr shifts, we measured the transmission through the mode $j$ with the network vector analyzer with fixed excitation power, and varied the excitation power of the mode $k$ with an additional microwave source.
As the excitation power of the mode $j$ is kept fixed, $\bar{n}_j$ is constant and the self-Kerr effect only gives a global offset.
To convert the excitation power of the mode $k$ to the photon number, we used the value of the $\alpha_k$ that we previously calibrated with the self-Kerr shifts. 
The bare resonance frequencies $\omega'_j - 1/2 K_{jj} \bar{n}_j$ were also extracted from the self-Kerr measurement. The measurement of the cross-Kerr coefficients therefore does not contain free parameters.\\
The measurements and the predicted shifts are plotted in Fig.~\ref{fig:cK}. The vertical error bars correspond to the errors of the resonance fit whereas the horizontal error bars represent the estimated errors of the photon calibration. We observe frequency shifts that are of the same order as the shifts predicted by the theoretical model. The $K_{32}$ could not be extracted from our measurements. \\
The cross-Kerr shifts were measured with a fixed drive frequency for the mode $k$. Although the self-Kerr shift of the mode $k$ is smaller than the line width of the resonance, this can cause deviations of the linear dependence of the photon number with drive power (as the drive can become slightly off-resonant) and thus to deviation from linear frequency shift for the mode $j$. This could explain the curvature observed in some of the cross-Kerr measurements.
Errors in the chain parameters $C$ and $C_0$ affect self- and cross-Kerr coefficients in the same way. As our method of using the self-Kerr shift to calibrate the photon number and the measuring the cross-Kerr shift probes the ratio between self- and cross-Kerr coefficients our result is only weakly affected by such errors.
We have also investigated the effect of disorder on the mode structure of the chain (not shown) \cite{Weissl}. The low frequency modes turn out to be robust against disorder as these long wave length modes average over a large number of junctions. For the same reasons also their Kerr-coefficients are immune to disorder.

\section{Summary \& Conclusion}

We have investigated extended plasma resonances in chains of Josephson junctions. We have measured a SQUID chain containing 200 junctions and were able to measure the 14 lowest resonant modes of the chain in a two-tone scheme. The observed resonance frequencies are in excellent agreement with the theoretical prediction that we obtained by diagonalizing the Hamiltonian in the linear limit.
Three of these plasma modes lie within our measurement band and can be accessed directly in the experiment.\\
We have presented a derivation of the Kerr-coefficients of the extended plasma resonances of Josephson junction chains. The non-linearity of the Josephson inductance was therefore treated in perturbation theory.
To compare the theoretically predicted frequency shifts to the shifts observed in the experiment, we use the self-Kerr effect to calibrate the photon numbers of the plasma modes as a function of the drive power. Then we compared the cross-Kerr shifts with the theoretical prediction and find reasonable agreement.\\
When studying the superconductor insulator transition in Josephson junction chains, the low frequency plasma modes constitute the electromagnetic environment of the junctions. This environment provides quantum fluctuations that enables quantum phase slips on the junctions. It is the interplay between the plasma modes and the quantum phase slips which depends on the ratio $E_J/E_0$, that characterizes the superconductor-insulator transition.\\
Here we have studied for the first time the non-linear interaction between the plasma modes. The signature of the superconducting-insulating transition on the microwave transmission spectrum of a JJ chain is currently under theoretical investigation. The theoretical description as well as the experiments presented in this work focus on uniform Josephson junction arrays but an extension to non-uniform arrays is straight forward. Indeed recent works \cite{Taguchi2015} discuss mode engineering of the resonant modes of a chain  by varying the junction parameters. This approach might be extended to also engineer the non-linear interaction between the modes and thus tailor an active medium for the use in travelling wave parametric amplifiers\cite{HoEom2012,OBrien2014}.

\acknowledgements
The authors thank A. Blais, I. M. Pop and N. Roch for fruitful discussions. T.W. and A.K.F. acknowledge support from the Grenoble Nanoscience Foundation, B.K. acknowledges support from SNF. F.H. and W.G. are supported by Institut universitaire de France. We also acknowledge support from the European Research council (grant no. 306731).

\bibliography{Kerr_coefficients}

\appendix

\section{Linear Hamiltonian, eigenvalues and eigenfunctions}
\label{App:A}
In this Appendix we give a rigorous derivation of the theoretical model presented above. Let us start from the Lagrangian Eq.~(\ref{eq:Lagmode1}) which can be written in the compact form
\begin{equation}
{\mathcal L}= \frac{1}{2}  \vec{\dot{\Phi}}^T  \widehat{C}  \vec{\dot{\Phi}} -  \frac{1}{2}  \vec{\Phi}^T  \widehat{L}^{-1} \vec{\Phi},
\end{equation}
with the flux vector
\begin{equation}
\vec{{\Phi}}^T  = \left({\Phi}_0, {\Phi}_1,\hdots , {\Phi}_N\right) ,
\end{equation}
and the capacitance and inverse inductance matrices defined as
\begin{equation}
\begin{split}
 \widehat{C} &= \left(
\begin{array}{cccccc} C+C_{in} & -C & 0 & \hdots & &\\
-C  &  C_0+2C & -C &  0 & \hdots &\\
0 & -C &  C_0+2C & -C &  0  & \hdots \\
\vdots & 0 & \ddots &  \ddots & \ddots & \ddots \\
& & & & & \\
\end{array}\right),
\label{eq:capmat}
\end{split}
\end{equation}
and
\begin{equation}
 \widehat{L}^{-1} = \left( \begin{array}{ccccc c} \frac{1}{L} & \frac{-1}{L} & 0 & \hdots & &\vspace{2mm}\\
\vspace{2mm} \frac{-1}{L} &  \frac{2}{L} & \frac{-1}{L} &  0 & \hdots &\\ \vspace{2mm}
0 & \frac{-1}{L} &  \frac{2}{L} & \frac{-1}{L} &  0  & \hdots \\ \vspace{2mm}
\vdots & 0 & \ddots &  \ddots & \ddots & \ddots \\ \vspace{2mm}
& & & & & \\
\end{array} \right).
\end{equation}

The momentum vector associated with the flux vector has the units of a charge and is given by
\begin{equation}
\vec{Q}=   \widehat{C}\vec{\dot{\Phi}}.
\end{equation}
With this the Legendre transformation gives the Hamiltonian (\ref{eq:modHam}).

In order to derive the eigenvalues and eigenfunctions of this Hamiltonian, let us write the total energy of the system as

\begin{equation}
E =  \frac{1}{2}  \dot{\vec{\phi}}^2 + \frac{1}{2}  \vec{\phi} \widehat{C}^{-1/2} \widehat{L}^{-1} \widehat{C}^{-1/2} \vec{\phi}
\label{eq:en}
\end{equation}
where
\begin{equation}
\vec{\phi}= \widehat{C}^{-1/2} \vec{\Phi}
\label{eq:phi}
\end{equation}
and $ \widehat{C}^{-1/2}  \widehat{C}^{-1/2} = \widehat{C}^{-1}$. A Fourier decomposition gives
\begin{equation}
 \vec{\phi}(t) = \sum_j  \vec{\psi}_{j} \left(\varphi^*_j e^{i\omega_jt} + \varphi_j e^{-i\omega_jt}\right).
\label{eq:Fi}
\end{equation}
$\vec{\psi}_{j}$ and $\omega_j$ satisfy the eigenvalue equation
\begin{equation}
\widehat{C}^{-1/2} \widehat{L}^{-1} \widehat{C}^{-1/2}  \vec{\psi_{j}} = \omega_j^2  \vec{\psi_{j}}.
\label{eq:eigp1}
\end{equation}
Inserting eqs. (\ref{eq:Fi}) and  (\ref{eq:eigp1}) back into eq. (\ref{eq:en}) we find
\begin{equation}
E = 2 \sum_j \omega_j^2 \vert \varphi_j \vert^2.
\label{eq:eigenval}
\end{equation}
The energy of the normal modes and thus also the Hamiltonian can be expressed in terms of creation and annihilation operators $\hat{H}_0= \sum_j \hbar\omega_j (\hat{a}_j^\dag \hat{a}_j +1/2)$ where
\begin{equation}
\varphi_j= \sqrt{\frac{\hbar}{2\omega_j}} \hat{a}_j.
\label{eq:op}
\end{equation}
The flux operator for the fluxes on the islands can be obtained by inverting Eq.~(\ref{eq:phi}) and inserting Eqs.~(\ref{eq:Fi}), (\ref{eq:eigp1}) and (\ref{eq:op})
\begin{equation}
\vec{\hat{\Phi}}=  \sum_{j}  \widehat{C}^{-1/2}\vec{\psi}_{j} \sqrt{\frac{\hbar}{2\omega_j}} \left( \hat{a}_j + \hat{a}_j^\dag\right) . \label{eq:Fluxop1}
\end{equation}

\section{Nonlinear Hamiltonian}
We include the lowest order nonlinear term of the Josephson energy, $-E_J\sum_{n=0}^{N-1}(\phi_{n+1} - \phi_n)^4/24$  as a perturbation to the linear Hamiltonian Eq.~(\ref{eq:modHam}). We expand this nonlinear term using the eigenfunctions Eq.~(\ref{eq:Fluxop1}). As a result we obtain the weak nonlinear contribution
\begin{equation}
\begin{split}
\hat{U}_{NL}= &-\frac{1}{24} \left(\frac{2 e}{\hbar}\right)^4 E_J  \sum_{n=0}^{N-1} (\Phi_{n+1} - \Phi_n)^4 \\
= &-\frac{1}{24} \frac{16\pi^4}{\Phi_0^4} E_J  \sum_{n=0}^{N-1}\left[ \sum_{m,j} \left(\widehat{C}_{n+1,m}^{-1/2}-\widehat{C}_{n,m}^{-1/2}\right)\psi_{m,j}\right. \\
&\ \ \ \ \ \ \ \ \left.\sqrt{\frac{\hbar}{2\omega_j}} \left( \hat{a}_j + \hat{a}_j^\dag\right) \right]^4\\
=&- \frac{\hbar^2 \pi^4 E_J }{6\Phi_0^{4} C^2} \sum_j \frac{ \eta_{jjjj}}{\omega_j^2} \left( \hat{a}_j + \hat{a}_j^\dag\right) ^4 \\
&- \frac{\hbar^2 \pi^4 E_J }{2\Phi_0^{4}C^2} \sum_{\begin{array}{c}j,k \\(j\neq m)\end{array}}\frac{ \eta_{jjkk}}{\omega_j\omega_k} \left( \hat{a}_j + \hat{a}_j^\dag\right) ^2\left( \hat{a}_k + \hat{a}_k^\dag\right) ^2\\
&- \frac{2\hbar^2 \pi^4 E_J }{3\Phi_0^{4}C^2} \sum_{\begin{array}{c}j,k \\(j\neq k)\end{array}}  \frac{ \eta_{jjjk}}{\omega_j\sqrt{\omega_j\omega_k}} \left( \hat{a}_j + \hat{a}_j^\dag\right) ^3\left( \hat{a}_k + \hat{a}_k^\dag\right) \\
&- \frac{\hbar^2 \pi^4 E_J }{\Phi_0^{4}C^2} \sum_{\begin{array}{c}j,k,l \\(j\neq k\neq l)\end{array}}  \frac{ \eta_{jjkl}}{\omega_j\sqrt{\omega_k\omega_l}} \left( \hat{a}_j + \hat{a}_j^\dag\right) ^2\\
& \ \ \ \ \ \ \ \ \ \ \ \ \  \ \ \ \ \ \ \ \ \ \ \ \ \ \ \ \ \ \ \ \ \times \left( \hat{a}_k + \hat{a}_k^\dag\right) \left( \hat{a}_l + \hat{a}_l^\dag\right) \\
&- \frac{\hbar^2 \pi^4 E_J }{6\Phi_0^{4}C^2} \sum_{\begin{array}{c}j,k,l,m \\(j\neq k\neq l\neq m)\end{array}}  \frac{\eta_{jklm}}{\sqrt{\omega_j\omega_k\omega_l\omega_m}} \left( \hat{a}_j + \hat{a}_j^\dag\right)\\
& \ \ \ \ \ \ \ \ \ \ \ \ \ \ \ \  \ \ \ \ \times \left( \hat{a}_k + \hat{a}_k^\dag\right) \left( \hat{a}_l + \hat{a}_l^\dag\right) \left( \hat{a}_m + \hat{a}_m^\dag\right) ,
\end{split}
\end{equation}
where we introduced the superconducting flux quantum $\Phi_0 = h/2 e$. Equation (\ref{eq:NLHam}) in the main text is then obtained by using
\begin{equation}
\left( \hat{a}_j + \hat{a}_j^\dag\right) ^4 = 6 \hat{a}_j^\dag\hat{a}_j\hat{a}_j^\dag\hat{a}_j + 6 \hat{a}_j^\dag\hat{a}_j +3,
\end{equation}
\begin{multline}
\left( \hat{a}_j + \hat{a}_j^\dag\right) ^2\left( \hat{a}_k + \hat{a}_k^\dag\right) ^2 \\
= 4  \hat{a}_j^\dag\hat{a}_j\hat{a}_k^\dag\hat{a}_k + 2 \hat{a}_j^\dag\hat{a}_j + 2 \hat{a}_k^\dag\hat{a}_k +1,
\end{multline}
\begin{multline}
\left( \hat{a}_j + \hat{a}_j^\dag\right) ^3\left( \hat{a}_k + \hat{a}_k^\dag\right) \\
= ( 3\hat{a}_j^\dag\hat{a}_j) (\hat{a}_j^\dag\hat{a}_k +  \hat{a}_k^\dag\hat{a}_j ) + 3\hat{a}_k^\dag\hat{a}_j,
\end{multline}
\begin{multline}
\left( \hat{a}_j + \hat{a}_j^\dag\right) ^2\left( \hat{a}_k + \hat{a}_k^\dag\right) \left( \hat{a}_l + \hat{a}_l^\dag\right)  \\
= ( 2\hat{a}_j^\dag\hat{a}_j+1) (\hat{a}_k^\dag\hat{a}_l +  \hat{a}_l^\dag\hat{a}_k ),
\end{multline}
and
\begin{multline}
\left( \hat{a}_j + \hat{a}_j^\dag\right)\left( \hat{a}_k + \hat{a}_k^\dag\right)\\
\left( \hat{a}_l + \hat{a}_l^\dag\right) \left( \hat{a}_m + \hat{a}_m^\dag\right)=\hat{a}_j^\dag \hat{a}_k \hat{a}_l^\dag\hat{a}_m + ....
\end{multline}

\end{document}